\documentclass[12pt,onecolumn,manuscript, tighten,trackchanges]{aastex631}%[12pt,onecolumn,manuscript,tighten,trackchanges,linenumbers]{aastex62}% % ,twocolumn,preprint % ,onecolumn,manuscript
\usepackage{natbib}%[comma]
\usepackage{multirow,color}
\usepackage{bbding}
\usepackage{amssymb}
\usepackage{ulem}
\usepackage{color}
\usepackage{longtable}
\usepackage{array}
\usepackage{bm}
\usepackage{url}
\usepackage{hyperref}
\usepackage{makecell}
\def\gsim{\;\lower4pt\hbox{${\buildrel\displaystyle >\over\sim}$}\;}
\def\lsim{\;\lower4pt\hbox{${\buildrel\displaystyle <\over\sim}$}\;}
\def\grls{\;\lower4pt\hbox{${\buildrel\displaystyle >\over <}$}\;}

\definecolor{darkcyan}{RGB}{0,100,230}%{0,102,204}

\newcommand{\q}{\color{red}} %\newcommand{\q}{\color{red}\bf}
\bibpunct{(}{)}{,}{o}{}{,} 

\begin{document}

\title{Photospheric horizontal magnetic field decrease preceding a major solar eruption}

\author{Lijuan Liu}
\affiliation{Planetary Environmental and Astrobiological Research Laboratory (PEARL), School of Atmospheric Sciences, Sun Yat-sen University, Zhuhai, Guangdong, 519082, China}
\affiliation{CAS center for Excellence in Comparative Planetology, China}
\affiliation{Key Laboratory of Tropical Atmosphere-Ocean System, Sun Yat-sen University, Ministry of Education, Zhuhai, China}
\author{Hanzhao Yang}
\affiliation{Planetary Environmental and Astrobiological Research Laboratory (PEARL), School of Atmospheric Sciences, Sun Yat-sen University, Zhuhai, Guangdong, 519082, China}

{\correspondingauthor{Lijuan Liu}}
{\email{liulj8@mail.sysu.edu.cn}}

%%%%%%%%%%%%%%%%%%%%%%%%%%%%%%%%%%%%%%%%%%%%%%%%%Abstract
\begin{abstract}

Significant photospheric magnetic field changes during major solar eruptions---interpreted as coronal feedback from eruptions to the photosphere---are well-observed. However, analogous short-time scale field changes preceding eruptions are rarely reported. 
In this study, 
we present the first detailed analysis %characterization 
of a pre-flare decrease in the photospheric horizontal magnetic field ($B_h$) associated with an X1.8 class flare, 
using high-cadence vector magnetic field data from Helioseismic and Magnetic Imager onboard Solar Dynamics Observatory (SDO). 
We identify a region of gradual, spatially coherent $B_h$ decrease of about 100 G along the flaring polarity inversion line (PIL) during 30 minutes preceding the flare. 
This decrease is accompanied by a decrease in the force-free parameter $\alpha_w$, with no significant flux emergence or cancellation observed. 
After the flare onset, $B_h$ exhibited contrasting behaviors in different sub-regions: a step-like increase near the PIL and a continued decrease in surrounding regions, 
suggesting that the pre-flare $B_h$ decrease may also have a coronal origin, like its post-flare counterparts. 
Coronal imaging from Atmospheric Imaging Assembly onboard SDO %SDO/AIA 
reveals that the associated erupting filament underwent a slow-rise phase before the flare, whose timing and location closely matches the occurrence of the pre-flare $B_h$ decrease. %EUV imaging
We propose that the slow-rise of the pre-eruptive filament 
stretched overlying coronal loops, increasing their verticality and thereby reducing $B_h$ at their photospheric footpoints. The results present the first detailed analysis of a pre-flare $B_h$ decrease and suggest it as a precursor to solar eruptions, causally linked to early filament activation and its impact on the photosphere.

\end{abstract}

%%%%%%%%%%%%%%%%%%%%%%%%%%%%%%%%%%%%%%%%%%%%%%%%Introduction
\section{Introduction}\label{sec:intro}

Solar active regions (ARs) which serve as the principal reservoir of major flares and coronal mass ejections (CMEs) harbor the strongest magnetic field in the solar atmosphere. 
Before an eruption, the gradual evolution of the ARs photospheric magnetic field, such as flux emergence, shearing motion, sunspot rotation, etc., helps to inject the excess magnetic energy into corona~\citep[e.g.,][]{Schrijver_2009, Wanghm_2015, Toriumi_2019}.  
When an eruption occurs, 
the magnetic energy stored in corona is released in an explosive manner 
through rapid reconfiguration of the coronal magnetic field, 
with part of the energy converting to the intense emission, and part converting to the kinematic energy of the accelerated particles and ejecta~\citep{Priest_2002a}. 
The pre-eruption evolution of the photospheric magnetic field usually occurs in a gentle manner, 
with a timescale ranging from hours to days, 
whereas the reconfiguration of coronal magnetic field related to eruptions occurs rapidly within seconds to minutes~\citep[e.g.,][]{Wangjx_2009b, Schrijver_2009}.

The timescale of the photospheric magnetic field evolution 
may be significantly shortened during major solar eruptions. %the 
Numerous observations have confirmed that remarkable changes in the photospheric magnetic field occur during eruptions~\citep[e.g.,][]{Wanghm_1994, Sudol_2005, Wangs_2012b, Duran_2017, Sun_2017, Petrie_2019, Liul_2022}. 
The most prominent change is the rapid and irreversible increase in the horizontal field component ($B_h$) that occurs below the post-flare loops~\citep[e.g.,][]{Wangs_2012b}. 
The increase has a magnitude typically on the order of hundreds Gauss~\citep[e.g.,][]{Liul_2022}. 
The evolution of the $B_h$ increase regions %regions where $B_h$ increase occurs 
has a close spatial-temporal relationship to the evolution of flare ribbons~\citep{Liuc_2018g,Liul_2022, Yadav_2023} as revealed by the high-cadence photospheric magnetograms provided by Helioseismic and Magnetic Imager~\citep[HMI;][]{Hoeksema_etal_2014, Sun_2017} onboard Solar Dynamics Observatory~\citep[SDO,][]{Pesnell_2012}.  
The observations, as well as numerical simulations~\citep{Barczynski_2019, Bian_2023}, suggest that $B_h$ increase is resulted from the contraction of post-flare loops, %and may also be 
and is related to magnetic implosion process from the perspective of energy release~\citep{Hudsons_2000, Yadav_2023}. 
Additionally, less prominent changes %, e.g., 
such as decrease of photospheric $B_h$ in the vicinity of $B_h$ increase region is frequently observed~\citep[e.g.,][]{Sun_2017}. 
This phenomenon is suggested to result from the straightening of the coronal loops stretched by the erupting CME~\citep{Barczynski_2019}.

The above observations reveal an 
inverse influence from the coronal activities to the photospheric magnetic field,  
which is counterintuitive given the dense nature of the photosphere~\citep{Aulanier_2016}. 
Apart from during the eruptions, analogous short time-scale changes in photospheric magnetic field 
preceding an eruption are much less frequently reported. %rarely reported. 
\citet{Murry_2012} documented an increase in vertical field strength on the order of hundreds Gauss and a change in the field inclination angle of $8^{\circ}$ prior to a GOES B1.0-class flare. Similarly, \citet{Maity_2024} showed %reported 
a pre-flare decrease in the horizontal field preceding to two M-class flares~\citep[Figure 3 in][]{Maity_2024}, though without %providing 
an explicit explanation. A pre-flare decrease in the vertical field strength by several tens of Gauss is also showed in \citet[][Figure 4]{Gong_2024} without detailed discussion. Besides changes in magnetic field, 
\citet{Liuy_2023} reported a pre-eruption variation in a photospheric magnetic parameter, $\alpha$, 
which quantifies the average twist of an AR~\citep[e.g.,][]{Seehafer_1990, Pevtsov_etal_1995}. 
Their results showed that $\alpha$ tends to decrease prior to the major eruption.

While the in-flare photospheric field changes 
are generally interpreted as a back-reaction of coronal eruptions, 
the mechanisms underlying the aforementioned pre-flare changes in photospheric magnetic field and related parameters remain poorly understood. %or
In particular, their potential correlation to the mild coronal or chromospheric activities such as precursor events preceding an eruption is still unclear. 
Precursors of an eruption can manifest in various forms in multiple passbands, 
including enhanced emission in multiple wavelengths~\citep[e.g.,][]{Asai_2006, Chifor_2007, Dudik_2016, Lid_2020}, 
chromospheric brightenings near the footpoints of sheared coronal loops~\citep{Wanghm_2017b}, 
coronal dimmings near the footpoints of pre-eruptive structures such as hot channels and filaments~\citep{Zhangqm_2017, Wangws_2023} which are seen as plasma proxy of magnetic flux ropes~\citep{Chengx_2017, Liur_2020a},  
broadening of the $H_{\alpha}$ line of pre-eruptive filaments~\citep{Chok_2016}, slow-rise of pre-eruptive hot channels or filaments~\citep{Zhang_2001a, Chengx_2020, Chengx_2023}, etc. 

Among those precursors,  
the slow-rise of pre-eruptive structures has been considered %in one study 
as a potential cause of the pre-eruption decrease in the photospheric force-free parameter $\alpha$~\citep{Liuy_2023}. 
\citet{Liuy_2023} suggested that the $\alpha$ decrease may be caused by either the slow-rise of a pre-eruptive magnetic flux rope which may lead to decrease in photospheric $B_h$ and consequently in the value of $\alpha$, or the emergence of a flux rope surrounded by a lower-twist region. However, a detailed analysis of these possibilities is lacking.       
Further study is needed to determine whether the precursor decrease of photospheric $B_h$ does occur accompanying the pre-flare decrease of $\alpha$, 
and if so, to explore its detailed evolution and potential correlation with the coronal precursor activities.

In this work, 
we identify a notable decrease of the photospheric horizontal field preceding an X1.8 class eruptive flare (SOL2011-09-07T22:32) and analyze it with the high-cadence magnetograms provided by HMI.  
The timing and location of the precursor decease of the horizontal field strongly suggests its correlation with the slow-rise of a pre-eruptive filament.

%%%%%%%%%%%%%%%%%%%%%%%%%%%%%%%%%%%%%%%%%%%%%%%%Method
\section{Data and event selection}\label{sec:data} 

The X1.8 class flare (SOL2011-09-07T22:32) analyzed in this study is selected from the flare list in~\citet{Liul_2022}, which studied the photospheric 
$B_h$ enhancement in 35 major solar flares with high-cadence HMI vector magnetic field data. 
Although not explicitly reported in \citet{Liul_2022}, we noticed that in a few cases the $B_h$ averaged within the post-flare $B_h$ enhancement region exhibited a slight decrease shortly before the flare onset in its temporal evolution curve.  
This study aims to investigate the detailed evolution and possible causes of this phenomenon. 
Considering that the pre-flare $B_h$ decrease, if confirmed, could originate either from the photospheric magnetic evolution (e.g., flux emergence/cancellation) or from coronal activities, particularly precursor activities of the flux rope---the core structure of solar eruptions, we search for events additionally having visible plasma proxy of the flux rope, such as a filament or a hot channel, from the flare list in~\citet{Liul_2022}. 

Furthermore, to examine the possible correlation between the short-term $B_h$ evolution and the pre-flare rise of the flux rope as suggested in~\cite{Liuy_2023}, it is necessary to measure the kinematics of the flux rope proxy. 
To enable reliable kinematic measurements, the event should not occur too close to the disk center. An eruption near the disk center introduces strong projection effects because it proceeds mainly along the line of sight (LOS), while the %SDO/AIA 
observations of Atmospheric Imaging Assembly instrument~\citep[AIA;][]{Lemen_etal_2012} onboard SDO provide only 2D motion in the plane of the sky (POS), perpendicular to the LOS. 
The 3D kinematics of filaments or hot channels can, in principle, be reconstructed using observations from multiple viewing angles, such as combined SDO and Solar Terrestrial Relations Observatory (STEREO) data~\citep[e.g., as in][]{Chengx_2020}. However, the STEREO viewing geometry is not always favorable, and the low-lying nature of some filaments or hot channels make their pre-flare 3D reconstruction difficult. 
Considering these factors, we selected this event as the optimal case for detailed investigation.

We use the data provided by HMI and AIA 
onboard SDO to do the analysis. HMI captures filtergrams in multiple polarization states at six wavelengths along the Fe {\scriptsize I} 6173~\AA~absorption line to extract the Stokes parameters, 
from which the photospheric vector magnetic field is derived 
using the Very Fast Inversion of the Stokes Vector algorithm~\citep{Hoeksema_etal_2014}. 
The field data has a plate scale of 0\farcs 5, and a cadence varying from 90 s to 720 s depending on the data product. 
In this study, we use the high-cadence data product \texttt{hmi.B\_135s}, which has a cadence of 135 s, 
to examine the photospheric field evolution 
preceding and throughout the flare. 
We create a set of cutout maps containing the source AR (NOAA AR 11283) of the flare from the full-disk magnetograms. 
For easier handling, we re-project the magnetic field vectors from the native Helioprojective-Cartesian coordinate to a local Cartesian cylindrical-equal-area (CEA) coordinate using the IDL procedure \texttt{bvec2cea.pro} available in Solar SoftWare (SSW) package~\citep{Sun_2013a}. The formal uncertainties of the data which is provided by the inversion code are likewise projected and resampled into the CEA coordinate using \texttt{bvecerr2cea.pro}.

To investigate the pre-flare coronal activities and the flare details, we use the Extreme Ultraviolet (EUV) images acquired by AIA onboard SDO. 
The images have a plate scale of 0\farcs 6 and a cadence of 12 s. Three channels, AIA 94~\AA, 211~\AA, 304~\AA~are mainly used. Additionally, we check the CME association with the flare by inspecting the Solar and Heliospheric Observatory~\citep[SOHO,][]{Domingo_2000} Large Angle and Spectrometric Coronagraph (LASCO) CME catalog\footnote{\url{https://cdaw.gsfc.nasa.gov/CME_list/index.html}}.

%%%%%%%%%%%%%%%%%%%%%%%%%%%%%%%%%%%%%%%%%%%%%%%%Result
\section{Results}\label{sec:res} 

\subsection{Flare details}\label{subsec:flare}

\begin{figure*}
\begin{center}
\epsscale{1.18}%0.85
\plotone{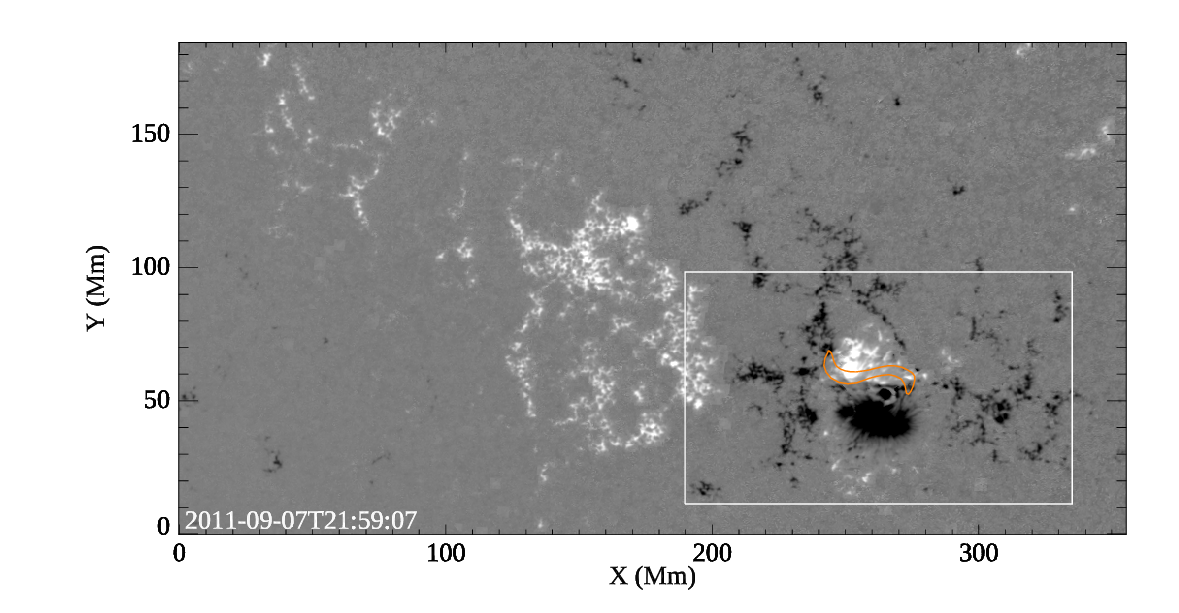} 
\caption{
Photospheric $B_z$ (vertical component of the magnetic field) map for NOAA AR 11283. 
The positive and negative magnetic field are shown in white and black, respectively, 
saturating at $\pm 1000$ Gauss. The white box indicates the field of view (FOV) of Figure~\ref{fig:bhd}. The orange line %delineates the skeleton of 
outlines the filament shown in Figure~\ref{fig:eru}(b). 
}\label{fig:bh_fila} 
\end{center}
\end{figure*}

\begin{figure*}
\begin{center}
\begin{interactive}{animation}{mov2.mp4}
\includegraphics[width=0.98\hsize]{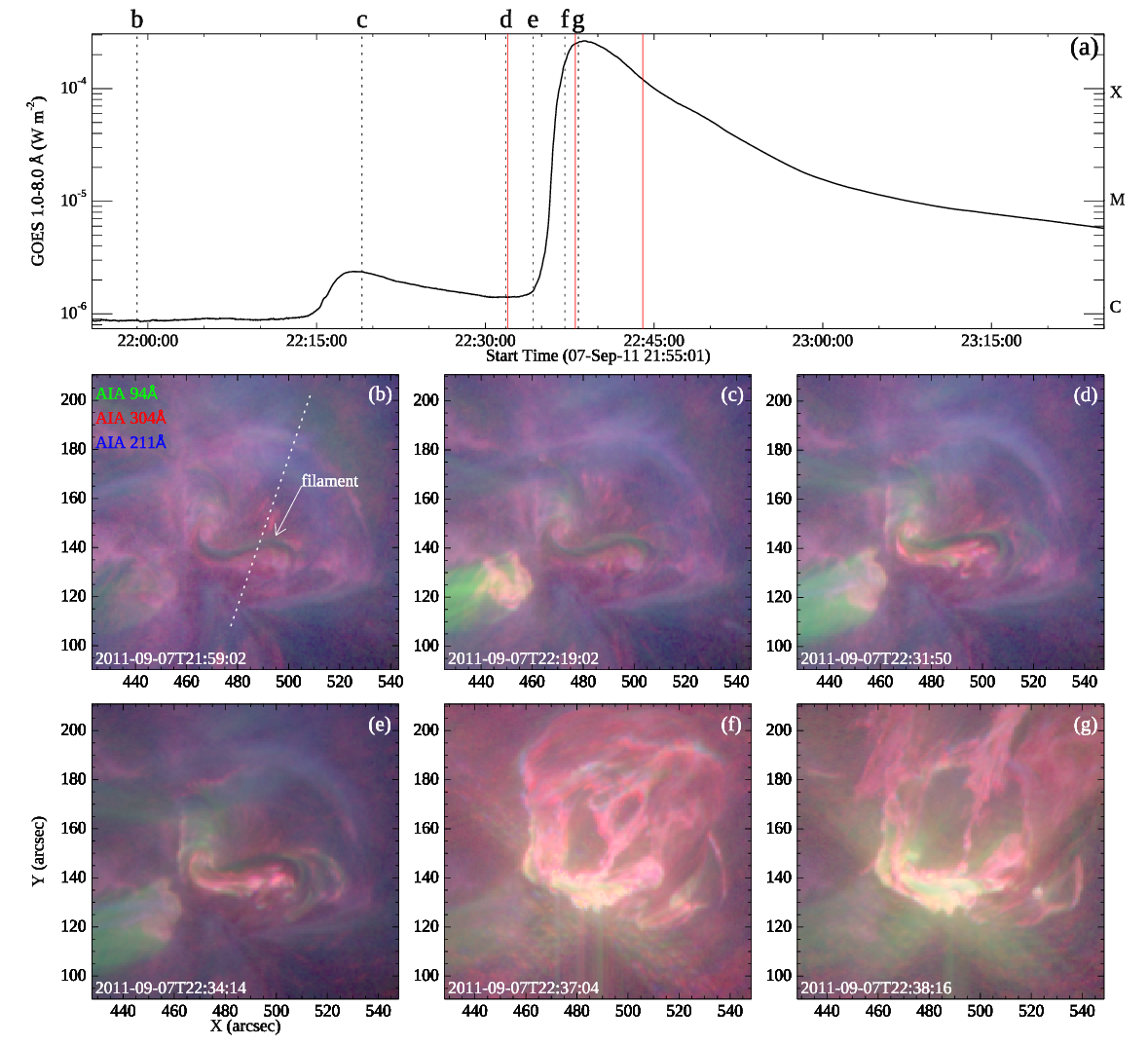} %0.72
\end{interactive}
%\epsscale{1.2}%0.96
%\plotone{Fig1}
\caption{
The filament eruption in the flare. %associated with the X1.8 flare. 
(a) Light curve of GOES soft X-ray flux. Vertical red lines mark the flare onset, peak, and end, 
while black dashed lines indicate the timings of panels (b)-(g). 
(b)-(g) Composite images of observations in AIA 94 \AA~, 304 \AA~, and 211 \AA, with panels (b)-(d) showing the filament evolution before the flare, and panels (e)-(g) showing the filament eruption during the flare. The white dashed slice in panel (b) is used to trace the filament rise (see in Figure~\ref{fig:height}). 
An animation %lasting from 2011-09-07T21:55 UT to 2011-09-07T23:25 UT, 
showing the filament evolution from 2011-09-07T21:55 UT to 2011-09-07T23:25 UT is available online. 
}\label{fig:eru} 
\end{center}
\end{figure*}

The X1.8 class flare (SOL2011-09-07T22:32) originated from NOAA AR 11283. The AR was an overall bipolar AR that emerged into a pre-existing network region (Figure~\ref{fig:bh_fila}), 
and was located not far from the central meridian (Stonyhurst N14W32) when producing the flare. 
The flare started from 2011-09-07T22:32 UT, peaked at 2011-09-07T22:38 UT, and ended at 2011-09-07T22:44 UT as revealed by the GOES soft X-ray light curve (Figure~\ref{fig:eru}(a)). 

The flare was associated with the eruption of a filament (Figure~\ref{fig:eru}). 
Before the flare, the filament was visible in multiple wavelengths, e.g., 304~\AA, 211~\AA, and even in 94~\AA~which is more sensitive to the high-temperature plasma (Figure~\ref{fig:eru}(b)). 
The main body of the filament was inclined toward the northern positive polarity 
(see orange curve in Figure~\ref{fig:bh_fila}). 
During about half an hour before %preceding 
the flare, the filament exhibited a slight slow-rise, 
with bright points appearing below the filament (Figure~\ref{fig:eru}(b)-(d)). 
The rise of the filament sped up near the flare onset (Figure~\ref{fig:eru}(e)), turning into a drastic eruption toward northwest afterwards (Figure~\ref{fig:eru}(f)-(g) and associated movie). The detailed kinematics of the filament is shown in the following (Figure~\ref{fig:height}). 

Accompanied by the filament eruption, a partial-halo CME with a central position angle of 290$^{\circ}$ (with respect to solar north) and first appearing at 2011-09-07T23:06 UT is recorded by SOHO LASCO/C2 CME catalog. 
Its timing and location coincide with the flare, confirming the flare's eruptive nature.

\subsection{Evolution of photospheric $B_h$}\label{subsec:bh}

The evolution of the photospheric magnetic field revealed an intriguing phenomenon: $B_h$ in a structured region along the flaring polarity inversion line (PIL; cyan lines in Figure~\ref{fig:bhd}) showed a gradual decrease before the flare (blue regions enclosed by black contours in Figure~\ref{fig:bhd}(b)-(d)). 
We demarcate the region from a base-difference $B_h$ map smoothed by the Fast Fourier Transform (FFT) method. 
The details of the method are presented in Appendix~\ref{app:ROI}. 
We refer this region as Region of Interest (ROI) in the following.

The majority of the region was located north of the PIL. The magnitude of $B_h$ decrease slowly intensified within about half an hour before the flare, with the mean value reaching around $-100$ Gauss. After the flare onset, a region of $B_h$ increase appeared around the PIL (red regions in Figure~\ref{fig:bhd}(d)), consistent with observations in other major eruptions~\citep{Liul_2022}. The region of post-flare $B_h$ increase was narrower, partially overlapping with the region of pre-flare $B_h$ decrease (ROI). Additionally, extensive, structured regions of $B_h$ decrease appeared on both sides of the post-flare $B_h$ increase region, with the northern region exhibiting much more pronounced changes (Figure~\ref{fig:bhd}(d)). The vertical magnetic field ($B_z$) showed no significant change and is not presented here. 

The high-cadence vector magnetic field data of HMI has larger uncertainties than lower-cadence (720 s) data~\citep{Sun_2017}. 
To assess the significance of the pre-flare $B_h$ decrease in this data, 
we compare the statistical characteristics of differenced $B_h$ in the ROI 
and that in a reference region outside the AR core 
(black box in Figure~\ref{fig:bhd}(c)). 
The reference region is chosen randomly. 
For all pixels in the two regions, we plot their distributions (Figure~\ref{fig:bhd}(e)-(f)) and calculate the median of differenced $B_h$ ($\tilde{\Delta B_h}$) %, 
along with its uncertainty to quantify the $B_h$ decrease. 
A Monte-Carlo experiment is performed to estimate the $1\sigma$ uncertainty, 
with the detailed procedure provided in Appendix~\ref{app:MC}. 
The uncertainties for the other parameters we analyzed (shown in Figure~\ref{fig:para}(b)-(f)) are similarly evaluated through an analogous Monte-Carlo process.

It is seen that the distributions of differenced $B_h$ for all pixels in the two regions are distinctly different (Figure~\ref{fig:bhd}(e)-(f)). 
Both distributions exhibit Gaussian-like shape. 
However, the distribution in the ROI is more skewed towards the negative side of differenced $B_h$, 
having a median value of $-92.6\pm 3.5$ Gauss, 
which is notably larger than that in the reference region of $-7.2\pm 2.1$ Gauss. 
The results confirm that pre-flare $B_h$ decrease in the ROI is statistically significant.

The temporal evolution of $B_h$ in the ROI is further investigated in detail (Figure~\ref{fig:para}).   
First, we check the unsigned magnetic flux (calculated by $\Phi=\Sigma |B_z|dA$) within the AR core region (Figure~\ref{fig:para}(a)). 
It is seen that except for a slight decrease of approximately $3\times 10^{20}$ Mx near the flare onset, the magnetic flux showed no significant variation, 
indicating that no substantial flux emergence or cancellation occurred during this stage. 

 We then analyze the evolution of the median $B_h$ within three small $3\times 3$ pixel boxes (each containing 9 pixels) centered on three selected pixels (yellow dots in Figure~\ref{fig:bhd}(c)), as shown in Figure~\ref{fig:para}(b)-(d). The pixels are chosen randomly from three representative regions:
pixels 1 and 2 are located within the ROI, with pixel 1 positioned closer to the PIL. Pixel 3 is located outside of the ROI and is selected for comparison. 
It is seen that median $B_h$ values in boxes surrounding pixel 1 and 2 both showed a pre-flare decrease for at least 30 minutes, 
with the decrease magnitude reaching $-163.8\pm 20.0$ Gauss ($1\sigma$ uncertainty estimated via the Monte-Carlo experiment; same in the following) for the former and $-136.6\pm 26.2$ Gauss for the latter. 
After the flare onset, their behavior displayed opposite evolutionary patterns: median $B_h$ in the box surrounding pixel 1 exhibited a step-wise increase, 
with the magnitude reaching $438.5\pm 37.2$ Gauss within 4.5 minutes, %463
whereas median $B_h$ in the box surrounding pixel 2 continued to decrease at a slighter quick rate compared to the pre-flare stage, reaching a magnitude of $-208.3\pm 31.9$ Gauss after 22.5 minutes. 
Median $B_h$ in the box surrounding pixel 3 exhibited no significant pre-flare change, 
and a slight decrease of $-30.1\pm 11.6$ Gauss within 16 minutes after the flare onset.

Furthermore, we check the evolution of median $B_h$ in the ROI (Figure~\ref{fig:para}(e)). 
The evolution of median $B_h$ reflected the collective behavior of most pixels within the ROI. 
It is seen that the median $B_h$ showed a continuous pre-flare decrease lasting for approximately 30 minutes, 
reaching a magnitude of $-104.7\pm 3.1$ Gauss. 
This was followed by an abrupt increase which commenced after the flare onset and reached a magnitude of $84.0\pm 3.6$ Gauss within 2 minutes. 
After the flare,  
the median $B_h$ remained a trend of very slight increase. 
We further perform a cubic spline smoothing to the median $B_h$ and calculate the first-order and second-order derivatives of smoothed $B_h$ over the time ($\displaystyle \frac{dB_h}{dt}$ and $\displaystyle \frac{d^2B_h}{dt^2}$; red and blue curves in Figure~\ref{fig:para}(e)) to quantify the median $B_h$ change. %the change of median $B_h$
The former reflects the changing rate of median $B_h$,  
while the latter indicates the ``acceleration'' of median $B_h$. 
From where $\displaystyle \frac{d^2B_h}{dt^2}=0$ we deduce that the systematical transition from $B_h$ decrease to $B_h$ increase within the ROI occurred $4.3$ minutes after the flare onset.

We also examine the evolution of the $B^2_z$-weighted, force-free parameter $\alpha_w$ within the ROI (Figure~\ref{fig:para}(f)). The $\alpha_w$ is calculate through %by the formula 
$\displaystyle \alpha_w=\frac{\int_{S} \alpha(x,y) B_z^2(x,y) dx dy}{\int_{S} B_z^2(x,y) dx dy}$, in which $\alpha(x,y)$ is calculated by $\alpha=[\nabla \times B]_z/B_z$ for each pixel~\citep{Hagino_2004, Liuy_2023}. It is seen that $\alpha_w$ did exhibit a pre-flare decrease, consistent with previous observations~\citep{Liuy_2023}. 
The decrease lasted for more than 30 minutes, 
reaching a magnitude of $-2.6\times 10^{-8}\pm 0.4\times 10^{-8}\ \mathrm{m^{-1}}$ and was followed by a rapid increase of $9.9\times 10^{-8} \pm 0.4\times 10^{-8}\ \mathrm{m^{-1}}$ within 7 minutes after the flare onset. 

\subsection{Kinematics of the filament}\label{subsec:filament}

Imaging observations suggest the filament underwent a slow-rise phase before the flare, 
and a main-acceleration phase after the flare onset (Figure~\ref{fig:eru}). 
To investigate the filament kinematics in detail, 
we first manually measure the projected height of the filament in the 304 \AA~images (Figure~\ref{fig:height}(a)), 
and smooth the height using the cubic spline method (green curves in Figure~\ref{fig:height}). 
The measured height confirms the filament experienced a slow-rise phase,  
rising by 3 Mm within approximately 30 minutes before the flare, 
and was followed by a drastic eruption associated with the main-acceleration phase after the flare onset. 

Previous studies suggest that the slow-rise phase could be fitted by a linear or quadratic function, 
while fitting of the main-acceleration phase requires a nonlinear function such as an exponential \citep[e.g.,][]{Chengx_2020}; 
the whole process can be fitted by a superposition of the two function forms~\citep[e.g.,][]{Chengx_2020}. 
We thus fit the filament rising with a prescribed function in the form of $\displaystyle h(t)=ae^{b(t-t_0)}+ct+h_0$, which is a combination of a linear and an exponential functions, 
using the IDL procedure {\it mpfit.pro}\footnote{\url{https://pages.physics.wisc.edu/~craigm/idl/fitting.html}}. 
The $a$, $b$, $t_0$, $c$, and $h_0$ are free parameters, 
with $t_0$ denoting the break point between the linear and exponential phases, 
$c$ denoting the velocity of the slow-rise phase, and $h_0$ denoting the initial height, respectively. 
The fitting yields that the filament rose at a velocity of around $1.1\pm 0.2\ \mathrm{km\ s^{-1}}$ in the slow-rise phase. The quoted uncertainty represents the $1\sigma$ error of the fitted parameter $c$ obtained from the fitting procedure. The breaking between the slow-rise and the main-acceleration phases occurred 2.5 minutes after the flare onset (indicated by a magenta vertical line in Figure~\ref{fig:height}(b)), slightly earlier than the systematical transition from $B_h$ decrease to increase ($4.3$ minutes after the flare onset). 

Note that the source AR was near the disk center (Stonyhurst coordinate N14W32) when the flare occurred, 
resulting in a considerable projection effect on the filament height measurement. 
This, in turn, led to a nontrivial underestimation of the filament height and velocity. 
We estimate how this projection effect influences the height and velocity measurement. The measured values represent only the components projected onto the POS. 
Assuming the eruption is radial, 
the true erupting height or velocity vector $\bm{v}$ 
is projected onto the POS as $\bm{v}_{proj}=\bm{v}\sqrt{1-cos\theta^2cos\phi^2}$, 
in which $\theta$ and $\phi$ are the latitude and longitude of the AR center. 
For $\theta=14^{\circ}$ and $\phi=32^{\circ}$ here, the projection (underestimation) factor $\sqrt{1-cos\theta^2cos\phi^2}$ is $0.57$, 
This implies corrected values of approximately $5.3$ Mm ($3/0.57$ Mm) for the rising height and $1.9\ \mathrm{km\ s^{-1}}$ ($1.1/0.57\ \mathrm{km\ s^{-1}}$) for the rising velocity during the slow-rise phase. The magnitude of the slow-rise velocity is comparable to that of the filament eruptions studied in~\citet[][F-label events in their Table 3]{Chengx_2020}. 

If the eruption is non-radial, the projection effect depends only on the angle between the eruption direction and the POS; a larger angle produces a stronger underestimation. In such case, a reliable correction is generally not feasible.
Nevertheless, the slow-rise phase of the filament prior to the flare is clearly discernible in this case.

%%%%%%%%%%%%%%%%%%%%%%%%%%%%%%%%%%%%%%%%%%%%%%%%Summary
\section{Summary and Discussion}\label{sec:sum} 

In this work, 
we identify a pre-flare $B_h$ decrease associated with an X1.8 class flare (SOL2011-09-07T22:32) and analyze it in detail using the SDO/HMI high-cadence vector magnetic field data. 
By employing the imaging observations provided by SDO/AIA, we further investigate the kinematics of the filament erupted in the flare 
to explore its potential correlation to the pre-flare $B_h$ decrease. 
The results are summarized below. 

\begin{itemize}

\item[1)]
  
$B_h$ in a structured region along the flaring PIL exhibited a gradual decrease by around $100$ Gauss within 30 minutes preceding the flare, 
being statistically significant compared to other regions. 
Most of the region located north of the PIL. 
Accompanied by the $B_h$ decrease, the force-free parameter $\alpha_w$ showed a pre-flare decrease as well, 
while the unsigned magnetic flux showed no significant change. 

\item[2)]

After the flare onset, 
an elongated region of $B_h$ increase appeared around the flaring PIL, 
surrounded by two extensive regions wherein $B_h$ showed decrease. 
Compared to the pre-flare $B_h$ decrease region (ROI), 
the post-flare $B_h$ increase region partially overlapped with the ROI, 
whereas the post-flare $B_h$ decrease regions were more extensive than the ROI. 
Accordingly, $B_h$ in different sub-regions comprising the ROI showed different post-flare evolutionary patterns, 
with $B_h$ closer to the PIL showing a step-wise increase by hundreds Gauss within a few minutes, while $B_h$ farther from the PIL continuing to decrease in a quicker rate for around 20 minutes.

\item[3)]

The filament associated with the flare was also inclined north of the PIL before the flare, 
and erupted northwestward %towards the northwest 
during the flare. 
Except the in-flare main-acceleration phase, 
the filament experienced a slow-rise phase preceding the flare.

\end{itemize}

The pre-flare $B_h$ decrease occurred in a structured region with dynamic evolution, and remained statistically significant despite high-noise of the data, indicating it has a physical origin rather than data noise. 
Moreover, in the sub-region of pre-flare decrease area located farther from the PIL, $B_h$ continued to decrease after the flare, suggesting a possible common physical origin for both pre-flare and post-flare $B_h$ decreases. 

While post-flare increase of $B_h$ is suggested to result from reconnection-driven contraction of post-flare loops~\citep{Barczynski_2019, Liul_2022, Bian_2023}, post-flare decrease of $B_h$ in the vicinity is attributed to coronal loop straightening caused by the erupting CME, which may lead to more vertical magnetic field and thereby reduced photospheric $B_h$~\citep{Barczynski_2019}. 
In our case, the filament erupted northwestward in the flare, and the northern part of the post-flare $B_h$ decrease regions showed more pronounced change, supporting that loop straightening driven by the eruption caused the post-flare $B_h$ decrease. 
During the filament slow-rise phase preceding the flare, 
although less violent than in the main-acceleration phase, it should more or less stretch the overlying coronal loops, 
resulting in reduced photospheric $B_h$ as well. 
Both the pre-flare filament and the pre-flare $B_h$ decrease region were inclined north of the PIL here, supporting a spatial correlation between them. 
Furthermore, the filament slow-rise and pre-flare $B_h$ decrease both began at least 30 minutes preceding the flare, supporting a temporal correlation between them as well. 
These observations together support that the pre-flare $B_h$ decrease is caused by the loop straightening driven by the slow rise of a pre-eruptive filament. 
It is reasonable that the post-flare $B_h$ decrease region is more extensive, 
given that the large-scale CME would involve more coronal loops than the pre-flare activity.

The above results present the first detailed analysis of the characteristics and evolution of a photospheric $B_h$ decrease before a major solar eruption, 
and suggest it is driven by the slow-rise of a pre-eruptive filament. 
This also provides an explanation to a pre-flare decrease of the force-free parameter $\alpha_w$, which is reported as a not uncommon phenomenon~\citep{Liuy_2023}. 
In our case, given the absence of significant flux emergence or cancellation during the pre-flare course,  
the $\alpha_w$ decrease is more likely linked to the $B_h$ change caused by the filament's slow-rise, rather than to the gradual evolution of photospheric magnetic field itself. 
The findings further imply that a few seemingly independent precursor events, 
including pre-flare decrease of $\alpha_w$~\citep{Liuy_2023}, 
pre-flare decrease of $B_h$ reported here, and coronal dimmings at the footpoints of pre-eruptive structures~\citep{Wangws_2023}, 
may all stem from the same underlying process: the slow-rise of pre-eruptive structures before solar eruptions. 
The slow-rise of the pre-eruptive structures can straighten the overlying coronal loops, 
leading to mass depletion and therefore dimming at the footpoints, and lowered $B_h$ on the photosphere.

Further statistical research is needed to assess whether the pre-flare $B_h$ decrease is a common feature of solar eruptions, and whether it is always related to the slow-rise phase of the erupting magnetic structure. In addition, 
its correlation with the flare magnitude, the flare eruptiveness, and the filament height and magnetic field strength at filament base all warrants further investigation. 
In our case, 
the low-lying filament may cause a stronger straightening effect on relatively lower coronal loops in its slow-rise phase, 
leading to a more pronounced imprint 
on the photospheric magnetic field. 
In contrast, high-lying filaments 
rooted in weaker photospheric magnetic field may produce less noticeable effects.

To answer above questions through statistical studies, high-quality photospheric vector magnetic field data together with EUV/ultraviolet (UV) imaging observations 
are required to analyze the pre-flare $B_h$ decrease and the pre-eruptive structures simultaneously. 
Events occurring neither too close to the disk center nor too close to the solar limb are preferred. 
Near-disk-center events suffer from strong projection effect when measuring the height of pre-eruptive filaments or hot channels, whereas events near limb have high noise level in the magnetic field measurements. 
Therefore, flares occurring between Stonyhurst longitudes W30$^{\circ}$-W60$^{\circ}$ and E60$^{\circ}$-E30$^{\circ}$ are good candidates. 
If the temporal evolution of $B_h$ is to be examined as well, high-cadence vector magnetic field data is needed. 
Above conditions suggest that one may begin with 
the flare list presented in~\citet{Liul_2022}.

In summary, the study reports the detailed characteristics and evolution of a pre-flare decrease in photospheric $B_h$ for the first time, 
and suggests it as a precursor to solar eruptions, causally linked to the slow-rise phase of the pre-eruptive filament. 

\clearpage

%%%%%%%%%%%%%%%%%%%%%%%%%%%%%%%%%%%%%%%%%%%%%%%%Figures

\clearpage

\clearpage

\begin{figure*}
\begin{center}
\begin{interactive}{animation}{mov3.mp4}
\includegraphics[width=0.98\hsize]{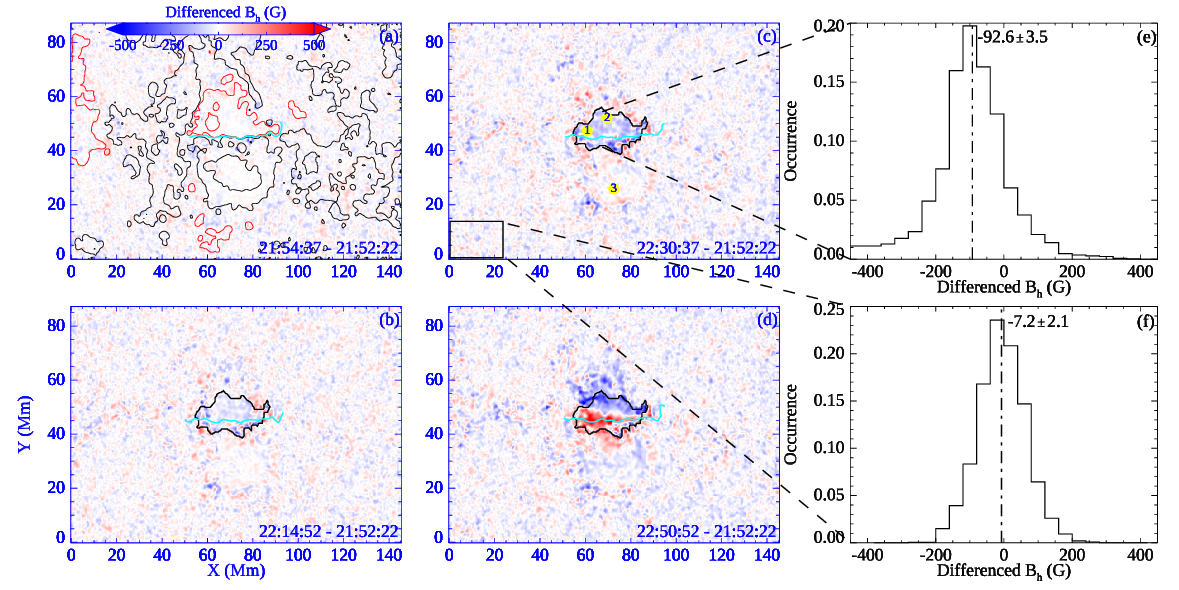} %0.68
\end{interactive}
%\epsscale{1.2}%0.96
%\plotone{Fig1}
\caption{
Evolution of photospheric $B_h$ in the AR core region (enclosed by white box in Figure~\ref{fig:bh_fila}). (a)-(d) Base-difference maps of $B_h$. 
Black (red) contours in panel (a) delineate the negative (positive) $B_z$ at -1000 and -100 (100 and 1000) Gauss, 
while cyan lines denote the flaring PIL. The black contours in panel (b)-(d) outline the pre-flare $B_h$ decrease region (ROI). 
The black box in panel (c) encloses the reference region to the ROI, 
and the yellow dots indicate pixels chosen %to show their $B_h$ evolution 
to show the evolution of median $B_h$ within three $9\times 9$ pixel boxes each centered on the selected pixels,
as plotted in Figure~\ref{fig:para}. 
(e)-(f) Distributions of differenced $B_h$ for all pixels within the ROI and the reference region, 
with vertical dashed lines marking their median values. 
The $1\sigma$ uncertainties of the median values are estimated via a Monte-Carlo experiment (see Appendix~\ref{app:MC} for details).
An animation showing $B_h$ evolution from 2011-09-07T21:54 UT to 2011-09-07T23:20 UT is available online. 
}\label{fig:bhd} 
\end{center}
\end{figure*}
\clearpage

\begin{figure*}
\begin{center}
\epsscale{1.18}% 0.95
\plotone{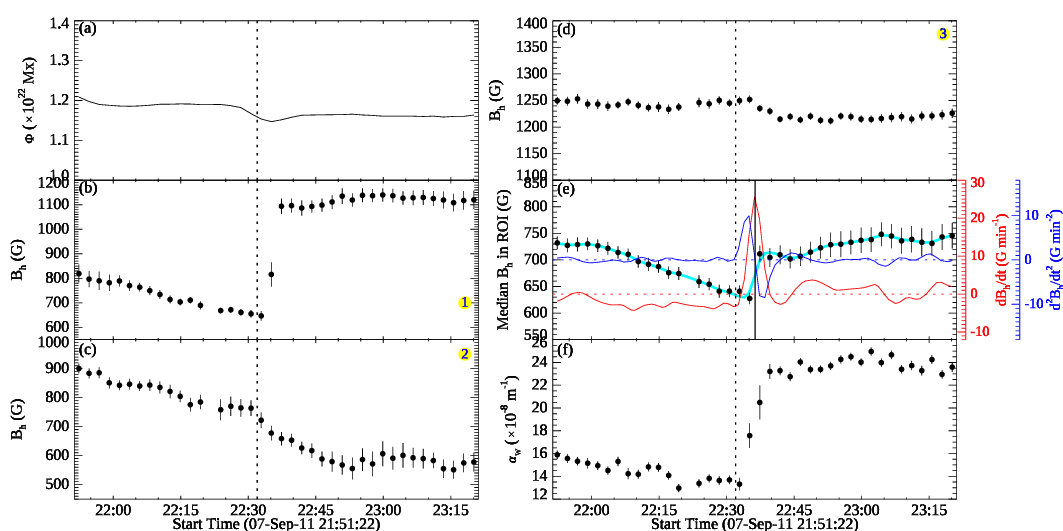}
\caption{Evolution of the parameters in the AR core region and the ROI. 
(a) Unsigned magnetic flux $\Phi$ in the AR core. 
The vertical dashed line marks the flare onset. %time (same in other panels). 
(b)-(d) %Evolution of $B_h$ in the pixels 1, 2 and 3 
Evolution of the median $B_h$ in three $9\times 9$ pixel boxes centered at the pixels 1, 2 and 3
(yellow dots in Figure~\ref{fig:bhd}(c)), with the first two in the ROI and the third outside the ROI. 
(e) Evolution of median $B_h$ in the ROI (black dots). 
The cyan curve represents the data smoothed using the cubic spline method. 
The red and blue curves show the first-order and second-order derivatives of smoothed $B_h$ over time ($\displaystyle \frac{dB_h}{dt}$ and $\displaystyle \frac{d^2B_h}{dt^2}$), respectively. 
The vertical solid line indicates the transition timing from $B_h$ decrease to increase deduced from where $\displaystyle \frac{d^2B_h}{dt^2}=0$. 
(f) Evolution of the force-free parameter $\alpha_w$ in the ROI. 
%Errors are propagated from the vector field data uncertainties. 
The error bars in Figure~\ref{fig:para}(b)–(f) represent $1\sigma$ uncertainties estimated using a Monte-Carlo experiment (for details see Appendix~\ref{app:MC}). 
}\label{fig:para} 
\end{center}
\end{figure*}
\clearpage

\begin{figure*}
\begin{center}
\epsscale{1.1}
\plotone{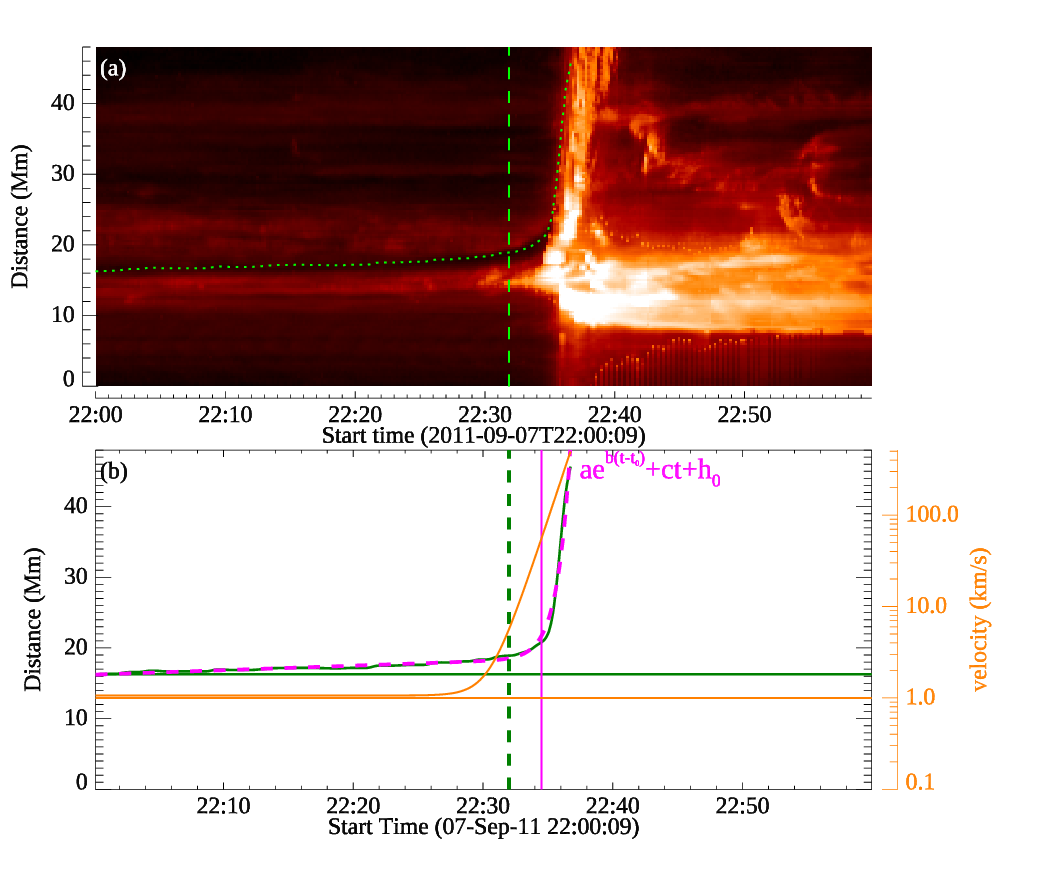}
\caption{
Details of the filament rising. (a) Time-distance diagram of the 304 \AA~intensity along the slice shown in Figure~\ref{fig:eru}(b). The green dashed curve depicts the smoothed projected height of the filament. 
The green vertical line marks the flare onset. 
(b) Kinematics of the rising filament. The green curve and the green vertical line %correspond to those in panel (a).
are the same as those shown in panel (a). The green horizontal line indicates the initial filament height (16.3 Mm from the slice start) as a reference. 
The magenta curve shows the fitting of the rising process, 
with the magenta vertical line indicating the break point between the linear and exponential lifting phases obtained from the fitting. 
%{\q The fitted parameters are $a=1.17\ Mm$, $b=0.96\ s^{-1}$, $t_0=2460.59\ s$, $c=0.001\ Mm\ s^{-1}$, and $h_0=15.8\ Mm$, respectively.} 
The yellow curve shows the velocity of the rising filament, 
with the yellow horizontal line marking the velocity of $1\ \mathrm{km\ s^{-1}}$ as a reference. 
%{\q The uncertainties of the measured height (each spanning 3 pixels, $\sim 1.34$ Mm) are assigned manually, while the velocity uncertainties are estimated by a Monte-Carlo experiment (see details in Appendix~\ref{app:vel}).} 
}\label{fig:height} %stack plot?
\end{center}
\end{figure*}
\clearpage

\appendix

\section{Identification of the pre-flare $B_h$ decrease region}\label{app:ROI} 

To identify the pre-flare $B_h$ decrease region (blue regions enclosed by black contours in Figure~\ref{fig:bhd}(b)-(d); referred to as ROI), 
we first construct the base-difference map by subtracting $B_h$ image at 2011-09-07T22:30 UT (right before the flare) from the $B_h$ image at 2011-09-07T21:52:22 UT (start time of the dataset), and then transform the data into the frequency domain using the FFT method. Second, we filter out the high-frequency components which are typically associated with the noise from the frequency domain. Third, we do a reverse FFT to transform the filtered data from the frequency domain back to the data domain. Finally, we grow the $B_h$ decrease region with a threshold of $-25$ Gauss in the FFT-smoothed $B_h$ difference map, and obtain the ROI.

\section{Uncertainty Estimation for the Pre-Flare $B_h$ Decrease and Related Parameters}\label{app:MC}
 
A Monte-Carlo experiment is performed to estimate the uncertainties of the pre-flare $B_h$ decrease and related parameters 
(shown in Section~\ref{subsec:bh}).  
Taking the median of $B_h$ decrease ($\tilde{\Delta B_h}$) in the ROI as an example (Figure~\ref{fig:bhd}(e)), 
for each pixel in the base-differenced $B_h$ map, subtracting pre-flare $B_h$ (2011-09-07T22:30 UT) from the value at the start of the dataset (2011-09-07T21:52:22 UT) introduces an uncertainty $\sigma_i$ of $\sqrt{B_{h,err1}^2+B_{h,err2}^2}$ according to standard error-propagation~\citep{Bevington_2003}, in which subscripts $1$ and $2$ denote the two measurement times. 

Using these propagated per-pixel uncertainties, we perform $R=2000$ Monte-Carlo realizations. 
In each realization, we add a random perturbation to every pixel draw from a normal distribution $\mathcal{N}(0,\sigma_i^2)$, 
and recompute $\tilde{\Delta {B_h}_{MC}}$ for the ROI in the perturbed map. 
After 2000 realizations, we obtain 2000 $\tilde{\Delta {B_h}_{MC}}$, 
and take their standard deviation $\sigma_{MC}$ as the Monte-Carlo estimate of the standard error of the original median differenced $B_h$ ($\tilde{\Delta B_h}$). 

At the meantime, we calculate the reduced Chi-square (with respect to the observed median $\tilde{\Delta B_{h}}$) in the ROI 
to assess whether the propagated per-pixel uncertainties underestimate the actual scatter in the data. 
The reduced Chi-square is calculated as $\displaystyle \chi_{\nu}^2=\frac{1}{\nu}\Sigma_i\frac{(\Delta B_{h,i}-\tilde{\Delta B_{h}})^2}{\sigma_i^2}$, in which $\Delta B_{h,i}$ and $\sigma_i$ are original base-differenced $B_h$ values and their propagated uncertainties for each pixel, 
and $\nu=N-1$ is the number of degrees of freedom for the total of $N$ pixels. 
We find $\chi_{\nu}^2>1$ in the ROI, indicating that the per-pixel uncertainties are underestimated by a factor of $\sqrt{\chi_{\nu}^2}$~\citep{Bevington_2003}.  
We therefore rescale 
the Monte-Carlo uncertainty as $\sigma_{\rm final}=\sqrt{\chi_{\nu}^2}\sigma_{MC}$. 
Finally, we report $\sigma_{\rm final}$ 
as the $1\sigma$ uncertainty for $\tilde{\Delta B_h}$. 

A similar Monte-Carlo procedure is applied to quantify the uncertainty of $\tilde{\Delta B_h}$ in the reference region (Figure~\ref{fig:bhd}(f)), to estimate the uncertainties of the median $B_h$ 
in the ROI and in the three $9\times9$ pixel boxes throughout their temporal evolution  (Figure~\ref{fig:para}(b)-(e)), and to evaluate the uncertainty in the temporal evolution of $\alpha_w$ (Figure~\ref{fig:para}(f)).

\begin{acknowledgments}
We thank our anonymous referee for the constructive comments that significantly improved the manuscript. 
We acknowledge the SDO, SOHO, and GOES missions for providing quality observations. 
Lijuan Liu acknowledges the support received from the National Natural Science Foundation of China (NSFC grant no. 12273123), 
and from the Guangdong Basic and Applied Basic Research Foundation (2023A1515030185).
\end{acknowledgments}

%\clearpage

\bibliographystyle{aasjournal} 
\bibliography{Pre_flare_BhD}

@article{Liul_2022,
	title = {On the {Nature} of the {Photospheric} {Horizontal} {Magnetic} {Field} {Increase} in {Major} {Solar} {Flares}},
	volume = {934},
	copyright = {All rights reserved},
	issn = {2041-8205},
	url = {https://iopscience.iop.org/article/10.3847/2041-8213/ac83bf},
	doi = {10.3847/2041-8213/ac83bf},
	number = {2},
	journal = {ApJL},
	author = {Liu, Lijuan and Zhou, Zhenjun and Wang, Yuming and Sun, Xudong and Wang, Guoqiang},
	month = aug,
	year = {2022},
	pages = {L33},
}

@article{Wanghm_2015,
	title = {Structure and evolution of magnetic fields associated with solar eruptions},
	volume = {15},
	issn = {1674-4527},
	url = {http://iopscience.iop.org/1674-4527/14/4/006},
	doi = {10.1088/1674-4527/15/2/001},
	number = {2},
	journal = {RAA},
	author = {Wang, Haimin and Liu, Chang},
	month = feb,
	year = {2015},
	note = {arXiv: 1311.6130},
	pages = {145--174},
}

@article{Barczynski_2019,
	title = {Flare {Reconnection}-driven {Magnetic} {Field} and {Lorentz} {Force} {Variations} at the {Sun}’s {Surface}},
	volume = {877},
	issn = {1538-4357},
	url = {http://dx.doi.org/10.3847/1538-4357/ab1b3d},
	doi = {10.3847/1538-4357/ab1b3d},
	number = {2},
	journal = {ApJ},
	author = {Barczynski, Krzysztof and Aulanier, Guillaume and Masson, Sophie and Wheatland, Michael S.},
	month = may,
	year = {2019},
	note = {arXiv: 1904.05447
Publisher: IOP Publishing},
	pages = {67},
}

@article{Liuc_2018g,
	title = {Evolution of {Photospheric} {Vector} {Magnetic} {Field} {Associated} with {Moving} {Flare} {Ribbons} as {Seen} by {GST}},
	volume = {869},
	issn = {1538-4357},
	url = {http://dx.doi.org/10.3847/1538-4357/aaecd0},
	doi = {10.3847/1538-4357/aaecd0},
	number = {1},
	journal = {ApJ},
	author = {Liu, Chang and Cao, Wenda and Chae, Jongchul and Ahn, Kwangsu and Choudhary, Debi Prasad and Lee, Jeongwoo and Liu, Rui and Deng, Na and Wang, Jiasheng and Wang, Haimin},
	month = dec,
	year = {2018},
	note = {Publisher: IOP Publishing},
	pages = {21},
}

@article{Sudol_2005,
	title = {Longitudinal {Magnetic} {Field} {Changes} {Accompanying} {Solar} {Flares}},
	volume = {635},
	issn = {0004-637X},
	doi = {10.1086/497361},
	number = {1},
	journal = {ApJ},
	author = {Sudol, J. J. and Harvey, J. W.},
	year = {2005},
	pages = {647--658},
}

@article{Wanghm_1994,
	title = {Vector magnetic field changes associated with {X}-class flares},
	volume = {424},
	journal = {ApJ},
	author = {Wang, HaimingGz and Ewell M.W, JR and {H. Zirin}},
	year = {1994},
	pages = {436--443},
}

@article{Aulanier_2016,
	title = {Solar physics: {When} the tail wags the dog},
	volume = {12},
	issn = {17452481},
	doi = {10.1038/nphys3938},
	number = {11},
	journal = {NatPh},
	author = {Aulanier, Guillaume},
	year = {2016},
	note = {Publisher: Nature Publishing Group},
	pages = {998--999},
}

@article{Toriumi_2019,
	title = {Flare-productive active regions},
	volume = {16},
	issn = {2367-3648},
	url = {https://doi.org/10.1007/s41116-019-0019-7},
	doi = {10.1007/s41116-019-0019-7},
	number = {1},
	journal = {LRSP},
	author = {Toriumi, Shin and Wang, Haimin},
	month = dec,
	year = {2019},
	note = {arXiv: 1904.12027
Publisher: Springer International Publishing
ISBN: 4111601900197},
	pages = {3},
}

@article{Pesnell_2012,
	title = {The {Solar} {Dynamics} {Observatory} ({SDO})},
	volume = {275},
	issn = {0038-0938},
	url = {http://link.springer.com/10.1007/s11207-011-9841-3},
	doi = {10.1007/s11207-011-9841-3},
	number = {1-2},
	journal = {SoPh},
	author = {Pesnell, W.{\textasciitilde}D. Dean and Thompson, B.{\textasciitilde}J. J. and Chamberlin, P.{\textasciitilde}C. C.},
	month = jan,
	year = {2012},
	note = {ISBN: 9781461436720},
	pages = {3--15},
}

@article{Wanghm_2017b,
	title = {High-resolution observations of flare precursors in the low solar atmosphere},
	volume = {1},
	issn = {2397-3366},
	url = {http://www.nature.com/articles/s41550-017-0085},
	doi = {10.1038/s41550-017-0085},
	number = {5},
	journal = {NatAs},
	author = {Wang, Haimin and Liu, Chang and Ahn, Kwangsu and Xu, Yan and Jing, Ju and Deng, Na and Huang, Nengyi and Liu, Rui and Kusano, Kanya and Fleishman, Gregory D. and Gary, Dale E. and Cao, Wenda},
	month = mar,
	year = {2017},
	note = {arXiv: 1703.09866
Publisher: Macmillan Publishers Limited, part of Springer Nature.},
	pages = {0085},
}

@article{Domingo_2000,
	title = {The scientific payload of the space-based {Solar} and {Heliospheric} {Observatory} ({SOHO})},
	volume = {70},
	issn = {0038-6308},
	url = {http://link.springer.com/10.1007/BF00777835},
	doi = {10.1007/BF00777835},
	number = {1-2},
	journal = {SSRv},
	author = {Domingo, V and Fleck, B and Poland, A. I.},
	month = oct,
	year = {1994},
	pages = {7--12},
}

@article{Zhang_2001a,
	title = {On the {Temporal} {Relationship} between {Coronal} {Mass} {Ejections} and {Flares}},
	volume = {559},
	issn = {0004-637X},
	url = {http://stacks.iop.org/0004-637X/559/i=1/a=452},
	doi = {10.1086/322405},
	number = {1},
	journal = {ApJ},
	author = {Zhang, J and Dere, K. P. and Howard, R. a. and Kundu, M R and White, S M},
	month = sep,
	year = {2001},
	pages = {452--462},
}

@article{Schrijver_2009,
	title = {Driving major solar flares and eruptions: {A} review},
	volume = {43},
	issn = {02731177},
	url = {https://linkinghub.elsevier.com/retrieve/pii/S0273117708005942},
	doi = {10.1016/j.asr.2008.11.004},
	number = {5},
	journal = {AdSpR},
	author = {Schrijver, Carolus J.},
	month = mar,
	year = {2009},
	note = {Publisher: Elsevier},
	pages = {739--755},
}

@article{Lemen_etal_2012,
	title = {The {Atmospheric} {Imaging} {Assembly} ({AIA}) on the {Solar} {Dynamics} {Observatory} ({SDO})},
	volume = {1},
	number = {275},
	journal = {SoPh},
	author = {Lemen, James R and Akin, David J and Boerner, Paul F and Chou, Catherine and Drake, Jerry F and Duncan, Dexter W and Edwards, Christopher G and Friedlaender, Frank M and Heyman, Gary F and Hurlburt, Neal E and {others}},
	year = {2012},
	pages = {17--40},
}

@article{Petrie_2019,
	title = {Abrupt {Changes} in the {Photospheric} {Magnetic} {Field}, {Lorentz} {Force}, and {Magnetic} {Shear} during 15 {X}-class {Flares}},
	volume = {240},
	issn = {1538-4365},
	url = {http://stacks.iop.org/0067-0049/240/i=1/a=11?key=crossref.3f5a626e9cd43ecbcaae5c897ff039f0},
	doi = {10.3847/1538-4365/aaef2f},
	number = {1},
	journal = {ApJS},
	author = {Petrie, Gordon J. D.},
	year = {2019},
	note = {Publisher: IOP Publishing},
	pages = {11},
}

@article{Hudsons_2000,
	title = {Implosions in {Coronal} {Transients}},
	volume = {531},
	issn = {0004637X},
	url = {http://stacks.iop.org/1538-4357/531/i=1/a=L75},
	doi = {10.1086/312516},
	number = {1},
	journal = {ApJ},
	author = {Hudson, H S},
	month = mar,
	year = {2000},
	pages = {L75--L77},
}

@article{Duran_2017,
	title = {A {Statistical} {Study} of {Photospheric} {Magnetic} {Field} {Changes} {During} 75 {Solar} {Flares}},
	volume = {852},
	issn = {1538-4357},
	url = {http://arxiv.org/abs/1711.08631},
	doi = {10.3847/1538-4357/aa9d37},
	number = {1},
	journal = {ApJ},
	author = {Durán, J. Sebastián Castellanos and Kleint, Lucia and Calvo-Mozo, Benjamín},
	year = {2017},
	note = {arXiv: 1711.08631
Publisher: IOP Publishing},
	pages = {25},
}

@article{Sun_2017,
	title = {Investigating the {Magnetic} {Imprints} of {Major} {Solar} {Eruptions} with {SDO}/{HMI} {High}-cadence {Vector} {Magnetograms}},
	volume = {839},
	issn = {1538-4357},
	url = {http://stacks.iop.org/0004-637X/839/i=1/a=67?key=crossref.8abc639e2317e3925c96f87b573df57a},
	doi = {10.3847/1538-4357/aa69c1},
	number = {1},
	journal = {ApJ},
	author = {Sun, Xudong and Hoeksema, J. Todd and Liu, Yang and Kazachenko, Maria and Chen, Ruizhu},
	year = {2017},
	note = {arXiv: 1702.07338
Publisher: IOP Publishing},
	pages = {67},
}

@article{Wangs_2012b,
	title = {{RESPONSE} {OF} {THE} {PHOTOSPHERIC} {MAGNETIC} {FIELD} {TO} {THE} {X2}.2 {FLARE} {ON} 2011 {FEBRUARY} 15},
	volume = {745},
	issn = {2041-8205},
	url = {http://stacks.iop.org/2041-8205/745/i=2/a=L17?key=crossref.30c4425a2fad28fbecebb53e73a3fe32},
	doi = {10.1088/2041-8205/745/2/L17},
	number = {2},
	urldate = {2014-11-13},
	journal = {ApJL},
	author = {Wang, Shuo and Liu, Chang and Liu, Rui and Deng, Na and Liu, Yang and Wang, Haimin},
	month = feb,
	year = {2012},
	pages = {L17},
}

@misc{Sun_2013a,
	title = {On the {Coordinate} {System} of {Space}-{Weather} {HMI} {Active} {Region} {Patches} ({SHARPs}): {A} {Technical} {Note}},
	url = {http://arxiv.org/abs/1309.2392},
	author = {Sun, Xudong},
	month = sep,
	year = {2013},
	note = {arXiv: 1309.2392
Volume: 94305},
}

@article{Liur_2020a,
	title = {Magnetic flux ropes in the solar corona: structure and evolution toward eruption},
	volume = {20},
	issn = {1674-4527},
	url = {http://iopscience.iop.org/1674-4527/14/4/006},
	doi = {10.1088/1674-4527/20/10/165},
	number = {10},
	journal = {RAA},
	author = {Liu, Rui},
	month = oct,
	year = {2020},
	note = {arXiv: 2007.11363},
	pages = {165},
}

@article{Chengx_2020,
	title = {Initiation and {Early} {Kinematic} {Evolution} of {Solar} {Eruptions}},
	volume = {894},
	issn = {0004-637X, 1538-4357},
	url = {https://iopscience.iop.org/article/10.3847/1538-4357/ab886a},
	doi = {10.3847/1538-4357/ab886a},
	language = {en},
	number = {2},
	urldate = {2023-07-31},
	journal = {ApJ},
	author = {Cheng, X. and Zhang, J. and Kliem, B. and Török, T. and Xing, C. and Zhou, Z. J. and Inhester, B. and Ding, M. D.},
	month = may,
	year = {2020},
	pages = {85},
}

@article{Liuy_2023,
	title = {Changes of {Magnetic} {Energy} and {Helicity} in {Solar} {Active} {Regions} from {Major} {Flares}},
	volume = {942},
	issn = {0004-637X},
	url = {https://dx.doi.org/10.3847/1538-4357/aca3a6},
	doi = {10.3847/1538-4357/aca3a6},
	language = {en},
	number = {1},
	urldate = {2023-07-26},
	journal = {ApJ},
	author = {Liu, Yang and Welsch, Brian T. and Valori, Gherardo and Georgoulis, Manolis K. and Guo, Yang and Pariat, Etienne and Park, Sung-Hong and Thalmann, Julia K.},
	month = jan,
	year = {2023},
	note = {Publisher: The American Astronomical Society},
	pages = {27},
}

@article{Hoeksema_etal_2014,
	title = {The {Helioseismic} and {Magnetic} {Imager} ({HMI}) {Vector} {Magnetic} {Field} {Pipeline}: {Overview} and {Performance}},
	volume = {289},
	doi = {10.1007/s11207-014-0516-8},
	journal = {SoPh},
	author = {Hoeksema, J T and Liu, Yang and Hayashi, Keiji and Sun, Xudong},
	year = {2014},
	pages = {3483--3530},
}

@article{Chengx_2017,
	title = {Origin and structures of solar eruptions {I}: {Magnetic} flux rope},
	volume = {60},
	issn = {1674-7313},
	url = {http://link.springer.com/10.1007/s11430-017-9081-x},
	doi = {10.1007/s11430-017-9074-6},
	number = {8},
	journal = {ScChD},
	author = {Cheng, Xin and Guo, Yang and Ding, MingDe},
	month = aug,
	year = {2017},
	note = {arXiv: 1706.05769},
	pages = {1383--1407},
}

@article{Pevtsov_etal_1995,
	title = {Latitudinal variation of helicity of photospheric magnetic fields},
	volume = {440},
	issn = {0004-637X},
	doi = {10.1086/187773},
	journal = {ApJ},
	author = {Pevtsov, Alexei a. and Canfield, Richard C. and Metcalf, Thomas R.},
	year = {1995},
	pages = {109--112},
}

@article{Seehafer_1990,
	title = {Electric current helicity in the solar atmosphere},
	volume = {125},
	journal = {SoPh},
	author = {{N.seehafer}},
	year = {1990},
	pages = {219--232},
}

@article{Priest_2002a,
	title = {The magnetic nature of solar flares},
	volume = {10},
	issn = {09354956},
	doi = {10.1007/s001590100013},
	number = {4},
	journal = {A\&ARv},
	author = {Priest, E. R. and Forbes, T. G.},
	year = {2002},
	pages = {313--377},
}

@article{Wangjx_2009b,
	title = {Magnetic changes in the course of the {X7}.1 solar flare on 2005 {January} 20},
	volume = {690},
	issn = {15384357},
	doi = {10.1088/0004-637X/690/1/862},
	number = {1},
	journal = {ApJ},
	author = {Wang, Jingxiu and Zhao, Meng and Zhou, Guiping},
	year = {2009},
	pages = {862--874},
}

@article{Chok_2016,
	title = {Strong {Blue} {Asymmetry} in {H}\${\textbackslash}upalpha\${Line} as a {Preflare} {Activity}},
	volume = {291},
	issn = {1573-093X},
	url = {https://doi.org/10.1007/s11207-016-0963-5},
	doi = {10.1007/s11207-016-0963-5},
	language = {en},
	number = {8},
	urldate = {2023-12-01},
	journal = {SoPh},
	author = {Cho, Kyuhyoun and Lee, Jeongwoo and Chae, Jongchul and Wang, Haimin and Ahn, Kwangsu and Yang, Heesu and Lim, Eun-kyung and Maurya, Ram Ajor},
	month = oct,
	year = {2016},
	pages = {2391--2406},
}

@article{Chengx_2023,
	title = {Deciphering the {Slow}-rise {Precursor} of a {Major} {Coronal} {Mass} {Ejection}},
	volume = {954},
	issn = {2041-8205},
	url = {https://dx.doi.org/10.3847/2041-8213/acf3e4},
	doi = {10.3847/2041-8213/acf3e4},
	language = {en},
	number = {2},
	urldate = {2023-12-07},
	journal = {ApJL},
	author = {Cheng, X. and Xing, C. and Aulanier, G. and Solanki, S. K. and Peter, H. and Ding, M. D.},
	month = sep,
	year = {2023},
	note = {Publisher: The American Astronomical Society},
	pages = {L47},
}

@article{Yadav_2023,
	title = {A {Statistical} {Analysis} of {Magnetic} {Field} {Changes} in the {Photosphere} during {Solar} {Flares} {Using} {High}-cadence {Vector} {Magnetograms} and {Their} {Association} with {Flare} {Ribbons}},
	volume = {944},
	issn = {0004-637X, 1538-4357},
	url = {https://iopscience.iop.org/article/10.3847/1538-4357/acaa9d},
	doi = {10.3847/1538-4357/acaa9d},
	language = {en},
	number = {2},
	urldate = {2024-04-01},
	journal = {ApJ},
	author = {Yadav, Rahul and Kazachenko, M. D.},
	month = feb,
	year = {2023},
	pages = {215},
}

@article{Asai_2006,
	title = {Preflare {Nonthermal} {Emission} {Observed} in {Microwaves} and {Hard} {X}-{Rays}},
	volume = {58},
	issn = {0004-6264},
	url = {https://doi.org/10.1093/pasj/58.1.L1},
	doi = {10.1093/pasj/58.1.L1},
	number = {1},
	urldate = {2024-04-01},
	journal = {PASJ},
	author = {Asai, Ayumi and Nakajima, Hiroshi and Shimojo, Masumi and White, Stephen M. and Hudson, Hugh S. and Lin, Robert P.},
	month = feb,
	year = {2006},
	pages = {L1--L5},
}

@article{Chifor_2007,
	title = {X-ray precursors to flares and filament eruptions},
	volume = {472},
	copyright = {© ESO, 2007},
	issn = {0004-6361, 1432-0746},
	url = {https://www.aanda.org/articles/aa/abs/2007/36/aa7771-07/aa7771-07.html},
	doi = {10.1051/0004-6361:20077771},
	language = {en},
	number = {3},
	urldate = {2024-04-01},
	journal = {A\&A},
	author = {Chifor, C. and Tripathi, D. and Mason, H. E. and Dennis, B. R.},
	month = sep,
	year = {2007},
	note = {Number: 3
Publisher: EDP Sciences},
	pages = {967--979},
}

@article{Dudik_2016,
	title = {{SLIPPING} {MAGNETIC} {RECONNECTION}, {CHROMOSPHERIC} {EVAPORATION}, {IMPLOSION}, {AND} {PRECURSORS} {IN} {THE} 2014 {SEPTEMBER} 10 {X1}.6-{CLASS} {SOLAR} {FLARE}},
	volume = {823},
	issn = {0004-637X},
	url = {https://dx.doi.org/10.3847/0004-637X/823/1/41},
	doi = {10.3847/0004-637X/823/1/41},
	language = {en},
	number = {1},
	urldate = {2024-04-01},
	journal = {ApJ},
	author = {Dudík, Jaroslav and Polito, Vanessa and Janvier, Miho and Mulay, Sargam M. and Karlický, Marian and Aulanier, Guillaume and Zanna, Giulio Del and Dzifčáková, Elena and Mason, Helen E. and Schmieder, Brigitte},
	month = may,
	year = {2016},
	note = {Publisher: The American Astronomical Society},
	pages = {41},
}

@article{Zhangqm_2017,
	title = {Pre-flare coronal dimmings},
	volume = {598},
	copyright = {© ESO, 2017},
	issn = {0004-6361, 1432-0746},
	url = {https://www.aanda.org/articles/aa/abs/2017/02/aa29477-16/aa29477-16.html},
	doi = {10.1051/0004-6361/201629477},
	language = {en},
	urldate = {2024-04-01},
	journal = {A\&A},
	author = {Zhang, Q. M. and Su, Y. N. and Ji, H. S.},
	month = feb,
	year = {2017},
	note = {Publisher: EDP Sciences},
	pages = {A3},
}

@article{Wangws_2023,
	title = {Investigating {Pre}-eruptive {Magnetic} {Properties} at the {Footprints} of {Erupting} {Magnetic} {Flux} {Ropes}},
	volume = {943},
	issn = {0004-637X, 1538-4357},
	url = {https://iopscience.iop.org/article/10.3847/1538-4357/aca6e1},
	doi = {10.3847/1538-4357/aca6e1},
	language = {en},
	number = {2},
	urldate = {2024-04-01},
	journal = {ApJ},
	author = {Wang, Wensi and Qiu, Jiong and Liu, Rui and Zhu, Chunming and Yang, Kai E. and Hu, Qiang and Wang, Yuming},
	month = feb,
	year = {2023},
	pages = {80},
}

@article{Lid_2020,
	title = {Preflare very long-periodic pulsations observed in {Hα} emission before the onset of a solar flare},
	volume = {639},
	copyright = {© ESO 2020},
	issn = {0004-6361, 1432-0746},
	url = {https://www.aanda.org/articles/aa/abs/2020/07/aa38398-20/aa38398-20.html},
	doi = {10.1051/0004-6361/202038398},
	language = {en},
	urldate = {2024-04-01},
	journal = {A\&A},
	author = {Li, Dong and Feng, Song and Su, Wei and Huang, Yu},
	month = jul,
	year = {2020},
	note = {Publisher: EDP Sciences},
	pages = {L5},
}

@article{Hagino_2004,
	title = {Latitude {Variation} of {Helicity} in {Solar} {Active} {Regions}},
	volume = {56},
	issn = {0004-6264},
	url = {https://doi.org/10.1093/pasj/56.5.831},
	doi = {10.1093/pasj/56.5.831},
	number = {5},
	urldate = {2024-07-26},
	journal = {Publications of the Astronomical Society of Japan},
	author = {Hagino, Masaoki and Sakurai, Takashi},
	month = oct,
	year = {2004},
	pages = {831--843},
}

@article{Bian_2023,
	title = {{MHD} simulation of rapid change of photospheric magnetic field during solar eruption caused by magnetic reconnection},
	volume = {10},
	issn = {2296-987X},
	url = {https://www.frontiersin.org/journals/astronomy-and-space-sciences/articles/10.3389/fspas.2023.1097672/full},
	doi = {10.3389/fspas.2023.1097672},
	language = {English},
	urldate = {2024-08-10},
	journal = {Front. Astron. Space Sci.},
	author = {Bian, Xinkai and Jiang, Chaowei},
	month = jun,
	year = {2023},
	note = {Publisher: Frontiers},
}

@article{Maity_2024,
	title = {Changes in {Photospheric} {Lorentz} {Force} in {Eruptive} and {Confined} {Solar} {Flares}},
	volume = {962},
	issn = {0004-637X},
	url = {https://dx.doi.org/10.3847/1538-4357/ad13f0},
	doi = {10.3847/1538-4357/ad13f0},
	language = {en},
	number = {1},
	urldate = {2025-03-10},
	journal = {ApJ},
	author = {Maity, Samriddhi Sankar and Sarkar, Ranadeep and Chatterjee, Piyali and Srivastava, Nandita},
	month = feb,
	year = {2024},
	note = {Publisher: The American Astronomical Society},
	pages = {86},
}

@article{Murry_2012,
	title = {The {Evolution} of {Sunspot} {Magnetic} {Fields} {Associated} with a {Solar} {Flare}},
	volume = {277},
	issn = {1573-093X},
	url = {https://doi.org/10.1007/s11207-011-9796-4},
	doi = {10.1007/s11207-011-9796-4},
	language = {en},
	number = {1},
	urldate = {2025-04-03},
	journal = {Sol Phys},
	author = {Murray, Sophie A. and Bloomfield, D. Shaun and Gallagher, Peter T.},
	month = mar,
	year = {2012},
	pages = {45--57},
}

@article{Gong_2024,
	title = {Rearrangement of sunspot magnetic field caused by an {X1}.5 solar flare},
	volume = {530},
	copyright = {https://creativecommons.org/licenses/by/4.0/},
	issn = {0035-8711, 1365-2966},
	url = {https://academic.oup.com/mnras/article/530/4/3897/7658461},
	doi = {10.1093/mnras/stae1020},
	language = {en},
	number = {4},
	urldate = {2025-04-09},
	journal = {Monthly Notices of the Royal Astronomical Society},
	author = {Gong, Liufan and Yan, Xiaoli and Liang, Hongfei and Xue, Zhike and Wang, Jincheng and Yang, Liheng and Peng, Yang and Yang, Liping and Zhang, Xinsheng},
	month = may,
	year = {2024},
	pages = {3897--3905},
}

@book{Bevington_2003,
	address = {Boston, Mass.},
	edition = {3. ed., [Nachdr.]},
	title = {Data reduction and error analysis for the physical sciences},
	isbn = {978-0-07-247227-1 978-93-392-2120-1},
	language = {eng},
	publisher = {McGraw-Hill},
	author = {Bevington, Philip R. and Robinson, D. Keith},
	year = {2003},
}

%\begin{thebibliography}{}
%\expandafter\ifx\csname natexlab\endcsname\relax\def\natexlab#1{#1}\fi
%\providecommand{\url}[1]{\href{#1}{#1}}
%\providecommand{\dodoi}[1]{doi:~\href{http://doi.org/#1}{\nolinkurl{#1}}}
%\providecommand{\doeprint}[1]{\href{http://ascl.net/#1}{\nolinkurl{http://ascl.net/#1}}}
%\providecommand{\doarXiv}[1]{\href{https://arxiv.org/abs/#1}{\nolinkurl{https://arxiv.org/abs/#1}}}

%\bibitem[{Asai {et~al.}(2006)Asai, Nakajima, Shimojo, White, Hudson, \&
%  Lin}]{Asai_2006}
%Asai, A., Nakajima, H., Shimojo, M., {et~al.} 2006, PASJ, 58, L1,
%  \dodoi{10.1093/pasj/58.1.L1}

%\end{thebibliography}

\end{document}